\documentclass[aip, reprint]{revtex4-1}

\draft 
\usepackage{graphicx}
\usepackage{subfigure}
\usepackage{ulem}
\usepackage{color}
\usepackage{bm}
\usepackage{comment}

\definecolor{electricpurple}{rgb}{0.75, 0.0, 1.0}

\newcommand{\MM}[1]{\textcolor{black}{#1}} 

\begin{document}


\preprint{Submitted to Nature Physics}

\title{Experimental observation of topologically protected helical edge modes in Kagome elastic plates}



\author{M. Miniaci}
\affiliation{University of Le Havre, Laboratoire Ondes et Milieux Complexes, UMR CNRS 6294, 75 Rue Bellot, 76600 Le Havre, France}
\affiliation{School of Aerospace Engineering and School of Mechanical Engineering, Georgia Institute of Technology}


\author{R. K. Pal}%
\affiliation{School of Aerospace Engineering and School of Mechanical Engineering, Georgia Institute of Technology}

\author{B. Morvan}
\affiliation{University of Le Havre, Laboratoire Ondes et Milieux Complexes, UMR CNRS 6294, 75 Rue Bellot, 76600 Le Havre, France}

\author{M. Ruzzene}%
\affiliation{School of Aerospace Engineering and School of Mechanical Engineering, Georgia Institute of Technology}
\email{ruzzene@gatech.edu}


\date{\today}


\maketitle 


\textbf{
The investigation of topologically protected waves in classical media\cite{Huber_NatPhys_2016} has opened unique opportunities to achieve exotic properties like one-way phonon transport~\cite{He_NatPhys_2016}, protection from backscattering and immunity to imperfections~\cite{Mousavi_Nat_Comm_2015}. Contrary to acoustic~\cite{yang2015topological} and electromagnetic~\cite{lu2014topological} domains, their observation in elastic solids has so far been elusive due to the presence of both shear and longitudinal modes and their modal conversion at interfaces and free surfaces.
Here we report the experimental observation of topologically protected helical edge waves in elastic media. The considered structure consists of an elastic plate patterned according to a Kagome architecture with an accidental degeneracy of two Dirac cones induced by drilling through holes. The careful breaking of symmetries couples the corresponding elastic modes which effectively emulates spin orbital coupling in the quantum spin Hall effect~\cite{kane2005quantum, hasan2010colloquium}.
The results shed light on the topological properties of the proposed plate waveguide and opens avenues for the practical realization of compact, passive and cost-effective elastic topological waveguides.}

Two broad ways to achieve topologically protected waveguides in phononic medi \cite{susstrunk2016classification}
include active systems, capable of breaking the time-reversal ($\mathcal{T}$) symmetry and mimicking the quantum Hall effect~\cite{Khanikaev_NatComm_2015, Wang_PRL_2015, Salerno_PRB_2016, Swintek_JAP_2015, Fleury_Science_2014, Nash_PNAS_2015}, and systems comprising solely of passive components assembled to establish analogues to the quantum spin Hall effect (QSHE)~\cite{Susstrunk_Science_2015, Ningyuan_PRX_2015, Pal_NJP_2017,pal2016helical, Salerno_NJP_2017,brendel2017pseudomagnetic}. Key to the latter analogy is the nucleation of a double Dirac cone and the coupling  of two degenerate modes corresponding to distinct irreducible representations of the reciprocal lattice symmetry group characterized by a Dirac dispersion\cite{Mousavi_Nat_Comm_2015}. Thus, systems with multiple degrees of freedom allow multiple possibilities for emulating the QSHE. Elastic plates 
and their wave mechanics emerge as excellent candidates due to the presence of an infinite number of modes with distinct polarizations and coupled deformation mechanisms. However, although this physics is attractive in terms of quest for topological phases, the following  challenges arise:
(i) in three dimensional solid structures, only shear and longitudinal modes exist and the dispersion behaviour is linear. Introducing geometrical modifications, either by reducing a dimension to get a plate or by introducing periodic holes and inclusions in the solid, multiple effective modes (Lamb waves, flexural, torsional waves) with intricate mode shapes and nonlinear dispersion branches are induced;
(ii) the geometry gets complex as the dispersion bands are tailored to achieve specific wave phenomena and multiple modes exist at any frequency posing problems for opening an isolated band gap;
(iii) strong backscattering and mode conversion is promoted at both free boundaries (voids) and interface between distinct materials (inclusions) by the large contrast in acoustic impedance.
These challenges have limited the investigations of topological properties of elastic systems, which have been so far restricted to mechanical lattices of coupled rigid bodies~\cite{Susstrunk_Science_2015}, acoustic metamaterials with resonators governed by scalar equations~\cite{lu2017observation}, and to the analysis of Valley modes for flexural waves in plates~\cite{Vila_ArXiv_Maggio_2017}. 

\begin{figure}
\centering
\begin{minipage}[]{1\linewidth}
{\includegraphics[trim=3mm 82mm 2mm 30mm, clip=true, width=1\textwidth]{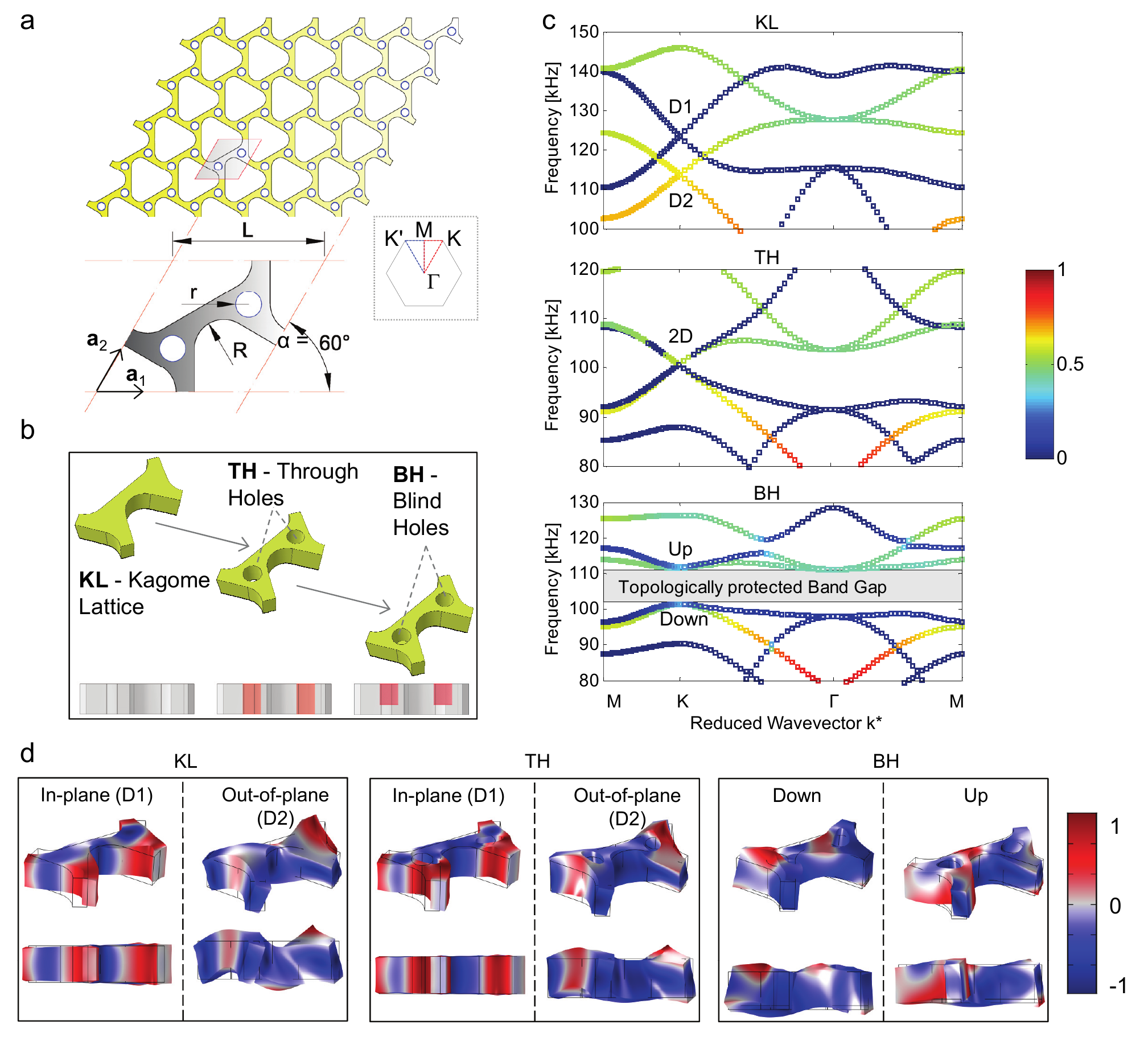}}
\end{minipage}
\caption{\textbf{Selective dispersion curve manipulation for the double Dirac cone nucleation and breaking of the $\sigma_{h}$ mirror symmetry for the opening of a topologically non-trivial band gap}. \textbf{a}, In-plane view of the phononic Kagome lattice and its unit cell schematic representation. The inset shows the $\Gamma$, $K$, $K'$ and $M$ high symmetry points of the first irreducible Brillouin zone, along which the dispersion curves are calculated. \textbf{b}, Perspective view of the unit cells used to selectively control dispersion branches: standard Kagome lattice (KL), Kagome lattice with holes drilled through the whole thickness (TH) and Kagome lattice with blind holes (BH). \textbf{c}, Phononic band structures for (upper panel) the KL, exhibiting two separate single Dirac points $D1$ and $D2$ at the $K$ point, (middle panel) the TH, allowing the formation of a double degenerate Dirac point $2D$ by inducing an accidental degeneracy in the band diagram and (lower panel) the BH, nucleating a topologically protected band gap ($9.7 \%$ relative width) by coupling of the $D1$ and $D2$ modes. Different mode polarizations are indicated by colors, ranging from pure in-plane, blue, to pure out-of-plane, red. \textbf{d}, Displacement fields at the $K$ point showing the deformation, the polarization and mode hybridization before, during and after the unit cell manipulation. Colours indicate the amplitude displacement.}
\label{fig1}
\end{figure}

Here we report an analogue of the QSHE in elastic plates. The first step in our process is to design a patterning on the plate such that the resulting periodic media band structure has two overlapping Dirac cones. The second step is  to open a topological band gap~\cite{Susstrunk_Science_2015,pal2016helical} by carefully breaking the relevant symmetries. In contrast to valley modes where the two sets of modes come from two opposite ($K$, $K^\prime$) valleys, here we work with two distinct sets of modes at each of the high symmetry points.
To nucleate a double Dirac cone, we consider an elastic plate patterned according to a Kagome lattice (KL) - Figs. \ref{fig1}a,b, since its band structure exhibits two distinct Dirac points (denoted as $D1$ and $D2$ in the top panel of Fig. \ref{fig1}c).
A similar configuration was recently investigated in a dual-scale phononic slab~\cite{Mousavi_Nat_Comm_2015} to achieve a double Dirac point. However this multi-scale approach drastically increased the engineering complexity due to a deep sub-wavelength patterning resulting in the bulk solid having a hexagonal ($C_{6}$ symmetry) and thus  yielding extreme elastic anisotropy. Here instead, we propose 
to work solely with isotropic bulk solids and selectively control specific dispersion branches by exploiting their sensitivity to geometric changes within the unit cell. We show how circular through the thickness holes (TH) arranged in a triangular fashion~\cite{Lu_PRB_2014} (Fig.~\ref{fig1}a,b) allow tailoring of the dispersion branches to obtain an isolated double Dirac point (named 2D in the middle panel of Fig. \ref{fig1}c). Finally, blind holes (BH) through part of the thickness of the plate breaks the $\sigma_h$ symmetry, while preserving the $C_{3v}$ symmetry to achieve coupling between the modes spanning the Dirac points, thereby emulating the spin orbital interaction in the QSHE~\cite{kane2005quantum, hasan2010colloquium}.


The dynamic response of the lattice is governed by the elastic equilibrium equation $\rho \ddot{\bm{u}} = (\lambda + \mu)\nabla (\nabla \cdot \bm{u}) + \nabla^2 \bm{u} = \bm{0} $ with $\rho$ being the density, $\bm{u}$ the displacement vector field and $\lambda,\mu$ the Lam\'{e} constants.
In the dispersion diagrams shown in Fig. \ref{fig1}c, colors indicate different mode polarization, ranging from pure in-plane (blue), to pure out-of-plane (red)~\cite{Miniaci_PRL_2017} - see Methods for metrics details. 
Note that both the Dirac points in the top panel of Fig. \ref{fig1}c correspond to essential degeneracies arising from the lattice having $D_{3h}$ symmetry. An examination of the mode shapes (left panel of Fig. \ref{fig1}d) reveals that the D1 and D2 modes span the subspaces associated respectively with the $E^{\prime}$ and $E^{\prime\prime}$ irreducible representations~\cite{dresselhaus2007group} of the reciprocal lattice group of wave vector $D_{3h}$ at the $K$ point. Furthermore, the mode shapes reveal that D1 and D2 are characterized by dissimilar displacement 
distributions. This implies that geometric modifications preserving lattice $D_{3h}$ symmetry, such as circular holes through the plate thickness (see TH geometry in Fig. \ref{fig1}b), will preserve the Dirac points, but produce different shifts in their frequency values. This simple approach allows to selectively shift the curves until the nucleation of a four-fold degeneracy with two overlaid Dirac cones (point 2D in the middle panel of Fig. \ref{fig1}c) is obtained (middle panel of Fig. \ref{fig1}c). Here, the double Dirac cone is achieved as an accidental degeneracy~\cite{huang2011dirac,chen2014accidental} by introducing through holes of radius $r = 0.085 L$, with $L = 20.5$ mm being the lattice parameter (see Methods and supplemental material - SM - for mechanical and geometrical details). 

Breaking the degeneracy at $K$  and coupling the D1 and D2 modes opens a topological band gap (low panel of Fig. \ref{fig1}c). Indeed, in contrast to through holes, which preserve the $\sigma_h$ symmetry without the occurrence of mode coupling, blind holes (BH geometry in Fig. \ref{fig1}b) couple the in-plane (D1) and out-of-plane (D2) modes, activating a mode hybridization process (Fig. \ref{fig1}d, right panel).
Breaking the $\sigma_h$ symmetry of the lattice lifts the degeneracy of both the D1 ($E^\prime$) and D2 ($E^{\prime\prime}$) modes. The hole depth which results in the accidental degeneracy between the two sets of separated modes is $0.9 H$, where $H$ is the plate thickness. This geometrical transformation converts the four-fold degeneracy into a $\approx 10 \%$-width non-trivial topologically protected band gap (lower panel of Fig. \ref{fig1}c). As the filled fraction in the hole decreases, the hybridization becomes stronger while the band gap becomes smaller (see SM). In addition, the absence of other modes at nearby frequencies, guarantees an isolated state for the hybridized modes. When two lattices with unit cells, related by $\sigma_h$ transformation, are joined together, topologically protected helical modes exist at the domain wall formed by the shared interface. Their existence is a consequence of the bulk boundary correspondence principle~\cite{hasan2010colloquium}, as the hybridized bulk modes on either side are distinct and related by a $\sigma_h$ transformation. In contrast, at the free boundary, these localized modes hybridize, become defect modes and do not span the bulk band gap~\cite{Mousavi_Nat_Comm_2015}.  

\begin{figure}
\centering
\begin{minipage}[]{1\linewidth}
{\includegraphics[trim=0mm 65mm 0mm 30mm, clip=true, width=1\textwidth]{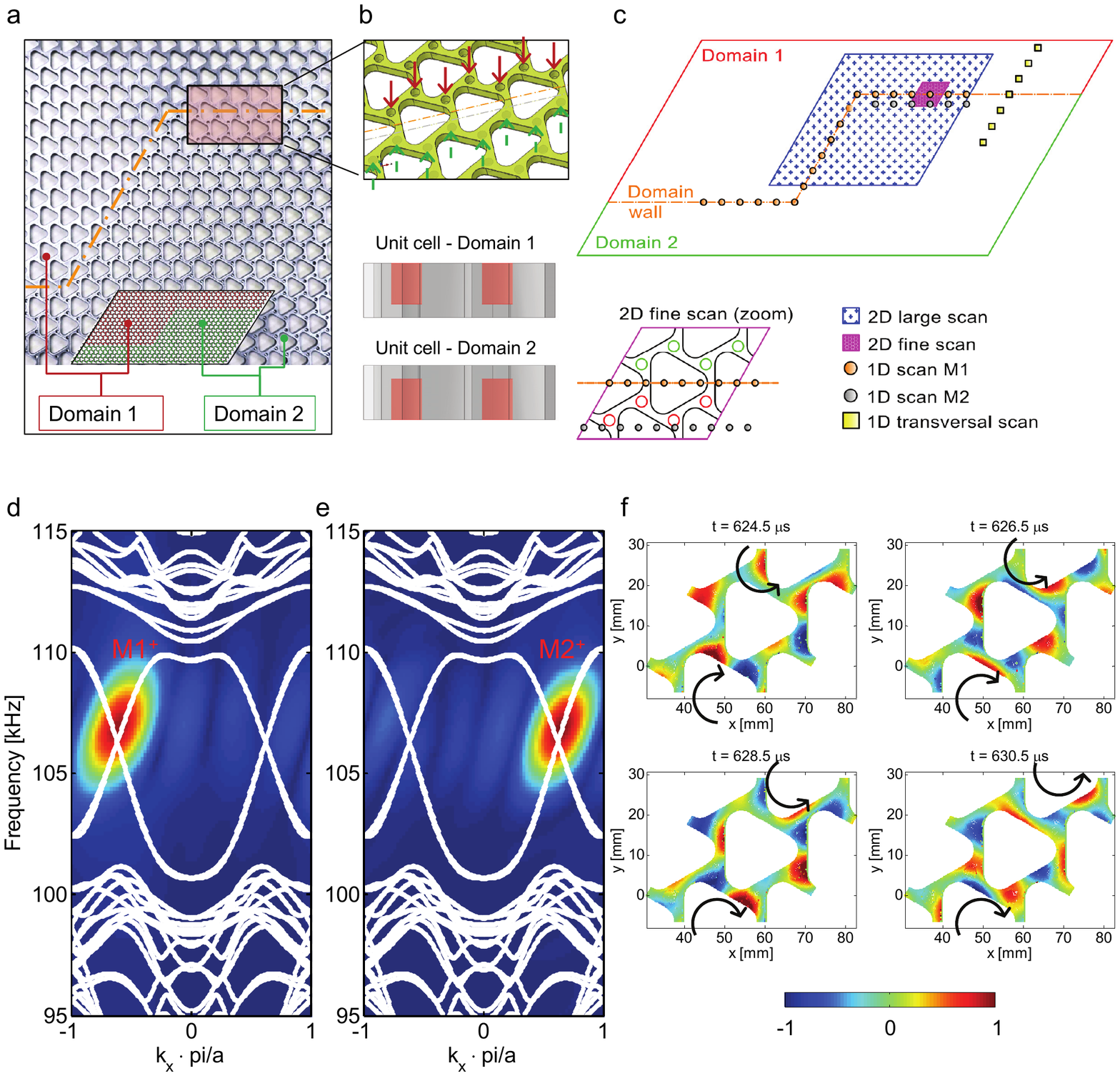}}
\end{minipage}
\caption{\textbf{Topologically protected waveguide supporting helical edge modes.}
\textbf{a}, The topological elastic waveguide consists of a periodic Kagome lattice with a Z-shaped domain wall (orange dashed line) separating two domains with blind holes drilled on opposite (top/bottom) surfaces of the plate. \MM{The inset shows the arrangement of the reversed unit cells on the full specimen (20 $\times$ 33 unit cells) divided into domain 1 in red and domain 2 in green.
\textbf{b}, A zoom of the domain wall and the cross-sections of the unit cells with the reversed holes. 
\textbf{c}, Synoptic picture of the scanned portions of the waveguide indicating the lines and areas of scan during the experimental measurements.
\textbf{d, e}, Measured dispersion curves for the right-propagating edge modes (denoted as M1$^+$ and M2$^+$, respectively) compared to the numerical predictions (white lines).
\textbf{f}, Measured field patterns for the right-propagating edge mode displaying typical vortex profiles. The black arrows indicate an anticlockwise and clockwise vorticity of the displacement field. The measure has been done along the area denoted as ``2D fine scan" in Fig. \ref{fig2}c.}}
\label{fig2}
\end{figure}

To confirm the emergence of topologically protected helical edge modes, we manufactured and investigated \MM{a plate with a Z-shape interface separating two domains with reversed blind holes (dotted orange line in Figs. \ref{fig2}a,b).} The Z-path is chosen to show the lack of back-scattering in the presence of sharp corners. \MM{A zoom of the domain wall and the arrangement of the reversed unit cells on the full specimen (made of 20 $\times$ 33 unit cells), divided into domain 1 (in red) and domain 2 (in green) are shown as well}. 

The dispersion curves calculated along a $10 \times 1$ periodic strip illustrate that the system exhibits a pair of helical edge modes (Figs. \ref{fig2}d,e) counter-propagating at the interface. \MM{The specimen shows an excellent transmission inside the gap (energy spots in Figs. \ref{fig2}d,e) due to edge modes hosted at the interface. They 
are labelled M1 and M2}, depending on their phase having an anticlockwise or clockwise polarization.

The frequency-wave number representation shown in Figs. \ref{fig2}d,e presents data for the normalized wave number $k_x  \pi/a $ from $-1$ to $+1$ and clearly identify the presence of the edge modes inside the gap, confirmed by the experimental energy distribution. The excitation frequency content inside the band gap is intentionally chosen to prevent bulk modes excitation. \MM{The two modes have been selectively detected by pointing the laser along specific 1D line scans of the unit cells composing the domain wall (see ``1D scan line M1" and ``1D scan line M2" in Fig. \ref{fig2}c).} Numerically predicted dispersion curves (the white lines) are superimposed onto the experimental data (obtained by plotting the maximum of the 2D-FFT of the acquired signals). An excellent agreement is found. Blue and red colours indicate the minimum and maximum Fourier amplitude. See Methods and SM for details.

To gain insight into the physics of these coupled modes, a fine scan ($0.2$ mm) on 4 unit cells is performed to experimentally reconstruct the full wavefield occurring during the propagation of the right-propagating edge mode, as presented in Fig. \ref{fig2}f (see the movie in the SM). Colours, varying from blue to red, indicate the out-of-plane displacement of the plate with respect to the unaltered configuration, respectively. The scan clearly shows the presence of both modes with vortex profiles. The black arrows indicate an anticlockwise or clockwise polarization associated with the phase of the displacement field. The vortex chirality can be viewed as a pseudo-spin analogous to the A-B sub-lattice or the top-bottom layer pseudo-spin in graphene systems~\cite{xiao2007valley}. 
The deformation mechanisms highlight how the system possesses invariant vortex cores with opposite chirality, as required by time-reversal symmetry which is indeed preserved in this approach. In contrast to the so far observed edge modes in acoustic/electromagnetic systems, where shear waves are not supported, here the information concerning the opposite chirality is brought by the out-of-plate thickness, which plays a key role in the reverse propagation process (Fig. \ref{fig2}f).

\begin{figure}
\centering
\begin{minipage}[]{1\linewidth}
{\includegraphics[trim=0mm 195mm 0mm 40mm, clip=true, width=1\textwidth]{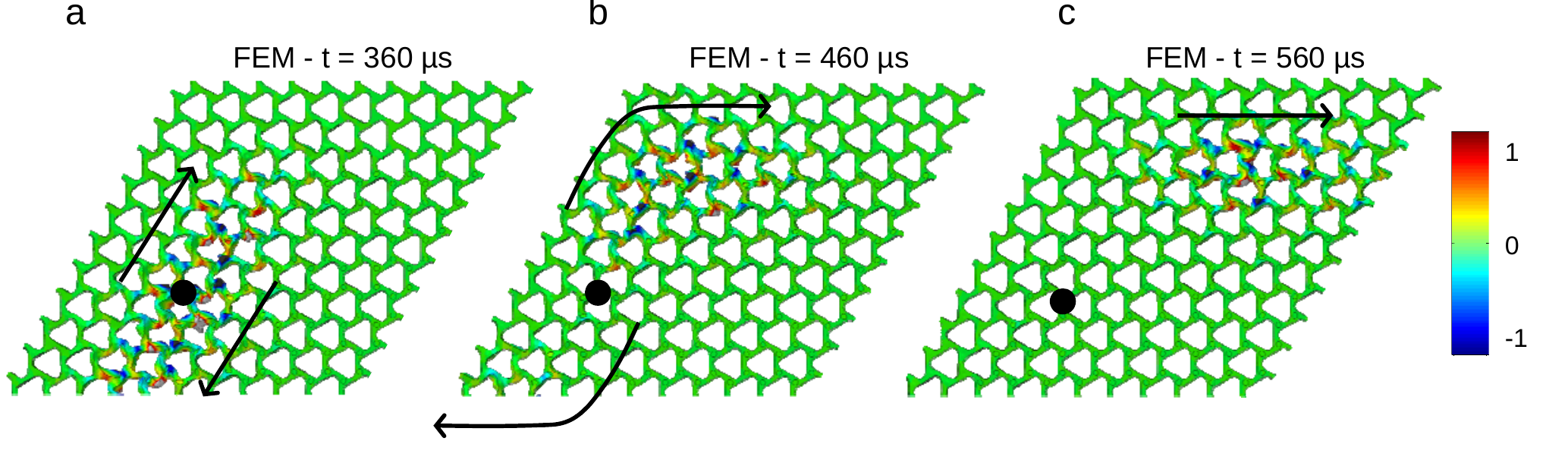}}
\end{minipage}
\caption{\textbf{Reflection immunity from a sharp Z-like pattern.} \textbf{a}, Transient dynamic simulated displacement field showing the excitation of the right-propagating and left-propagating helical edge modes. It is possible to notice the weak penetration of the waves inside the bulk region of the lattice. The black dot represents the excitation point. The deformation map is reconstructed after $t = 360 \mu$s. \textbf{b}, Evidence of how the non-trivial interface supports the energy transport beyond the $120^{\circ}$ bend without any localization or backscattering typical of trivial waveguides ($t = 460 \mu$s). \textbf{c}, Once the wave has passed the Z-bend it continues propagating in the right-direction ($t = 560 \mu$s).
Colours, varying from blue to red, indicate the displacement amplitude with respect to the undeformed configuration, respectively. Data are normalized to the maximum displacement value. Deformation is amplified by a factor of $1 \times 10^5$ to emphasize the displacement field. The analysed domain is the one denoted as "2D large scan" in Fig. \ref{fig2}c.}
\label{fig3}
\end{figure}

Finally, to verify the phenomenon of negligible backscattering in monolithic elastic waveguides, even in presence of a sharp bend ($120^{\circ}$) of the interface, a full 3D transient wave propagation event is first simulated and then experimentally measured on the manufactured structure. Fig. \ref{fig3} reports a map of the numerically predicted out-of-plane displacement distribution consequent to the excitation of elastic guided waves at the center of the structure (black dot) with a frequency content falling inside the bulk band gap. Wave-fields reconstructed before, while and after the edge mode is traversing the bend, i.e. after $t = 360 $ $\mu$s, $t = 460 $ $\mu$s and $t = 560 $ $\mu$s, respectively, clearly demonstrate  how the proposed approach allows a change in the direction of wave propagation (black arrows) without establishing standing wave patterns, typical of trivial waveguides. Besides the lack of backscattering, a weak penetration inside the bulk region is also noticeable.

\begin{figure}
\centering
\begin{minipage}[]{1\linewidth}
{\includegraphics[trim=0mm 130mm 0mm 50mm, clip=true, width=1\textwidth]{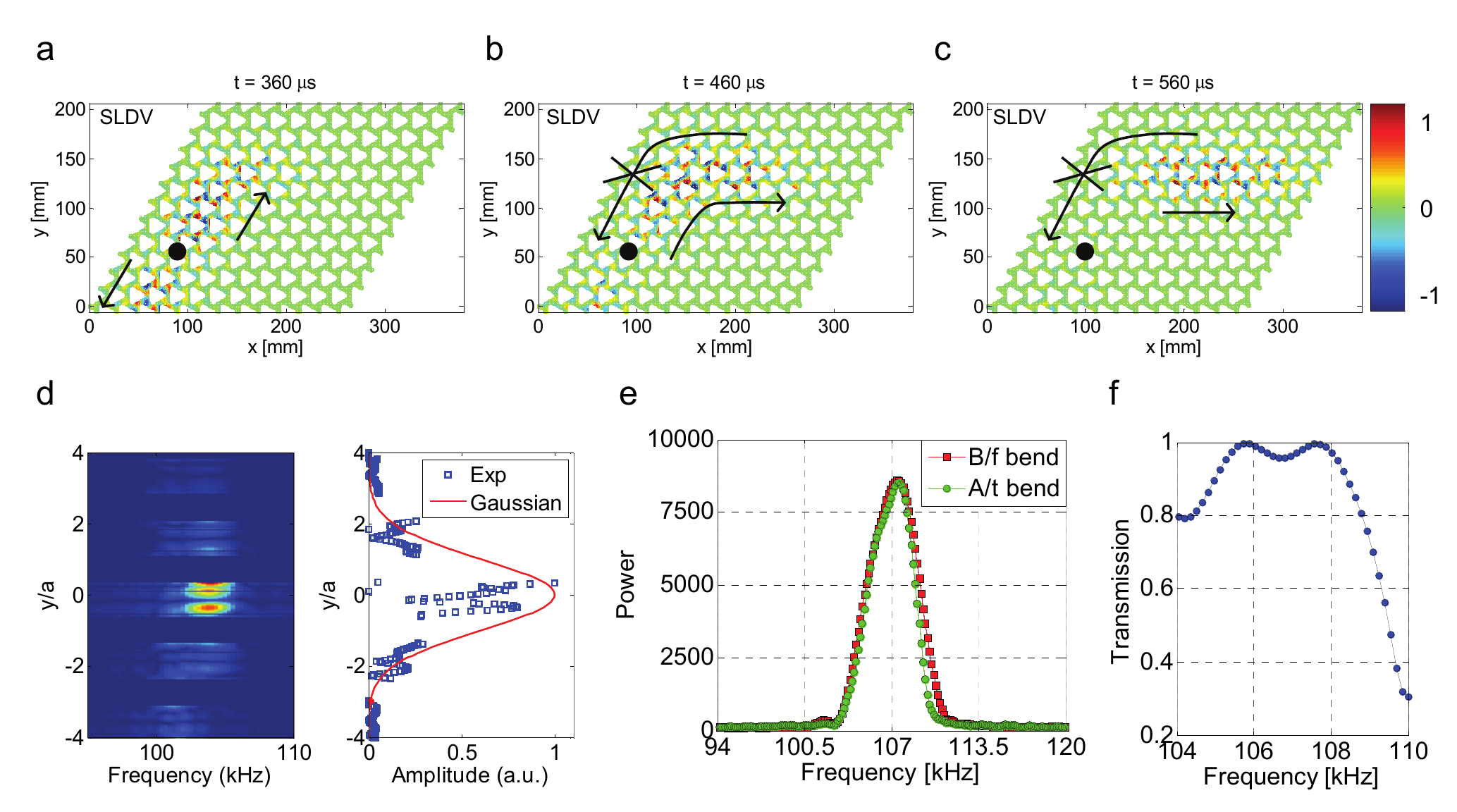}}
\end{minipage}
\caption{\textbf{Lack of backscattering and bulk penetration.} \textbf{a-c}, Experimental full wave field reconstruction of the right-propagating edge mode (in the area denoted as "2D large scan" in Fig. \ref{fig2}c). The black dot represents the excitation point and the arrows indicate the allowed and inhibited propagation directions. Colours, varying from blue to red, indicate the amplitude of the out-of-plane displacement of the plate. \textbf{d}, Experimental decay profile of the right-propagating edge mode away from the interface ($y = 0$) normalized by the lattice unit cell $a$: (left panel) a 2D-FFT representation as a function of the frequency and the normalized amplitude at the frequency of $106$ kHz. In the right panel a Gaussian is also reported as reference. Data are normalized by the corresponding maximum value. Signals are acquired along a line normal to the domain wall in the top part of the waveguide (see Fig. \ref{fig2}c) and similar results are observed for the clockwise polarized edge mode. \textbf{e}, \MM{The transmitted power of the propagating M1$^+$ mode as a function of the frequency. The energy is obtained by a 2D-FFT over scan lines of the same length placed before and after the upper sharp bend of the waveguide, quantitatively illustrating the absence of backscattering or mode conversion after the $120^{\circ}$ bend. \textbf{f}, Transmission coefficient for the mode M1$^+$ as a function of the frequency.}}
\label{fig4}
\end{figure}

The experimental full wavefield reconstruction by means of a scanning laser Doppler vibrometer in the top bend of the structure (similar physics is found in the lower bend) confirms the above presented picture. Indeed, the measured out-of-plane velocity distribution detected within \MM{the region labelled ``2D large scan" in Fig. \ref{fig2}c}, is in excellent agreement with the numerical predictions and clearly experimentally illustrates the topologically protected transportation properties of the proposed device. In both the simulations and experiments, the excitation frequency content was intentionally centred inside the bulk band gap to minimize  the excitation of bulk modes. This also allowed to quantitatively identify the exponential decay of the field amplitude away from the interface by scanning the deformation field along a straight line transversal to the domain wall (Fig. \ref{fig2}c). Plotting the maximum values of the 2D-FFT of the acquired signals allows to show the decaying path of the wave as a function of the frequency and the distance $y / a$ from the interface $y = 0$ (left panel of Fig. \ref{fig4}d), revealing its insensitivity to the frequency of excitation (right panel of Fig. \ref{fig4}d, showing the mentioned decaying profile at 106 kHz).

\MM{Figs. \ref{fig4}e shows the comparison of the energy content as a function of the frequency for the M1$^+$ mode before and after the sharp bend. A direct comparison is made possible because pointing the laser exactly along the domain wall ensures that we follow the evolution of a single mode, thereby allowing a quantitative evaluation of backscattering. Fig. \ref{fig4}e, obtained by 2D Fast Fourier transforming the data acquired along two 1D-line scans of the same length token before and after the bend, illustrates the insensitivity of bending around corners on the interface energy transport. In addition, Fig. \ref{fig4}f provides the transmission coefficient. It quantitatively proves that the change in the direction of the wave propagation (before/after bending) does not produce any significant energy lost, unambiguously proving that no mode conversion occurs at the bend, i.e. the left-propagation mode is inhibited.}

The results presented in this Letter clearly provide an experimental demonstration of topological robustness and the ability to guide waves along channels with sharp corners in elastic structures. Our design principle and measurement techniques, providing insights on the mechanism of band control to surpass the limits of conventional elastic waveguides, can be extended to explore waveguiding along arbitrary shaped pathways. The results may open up new avenues \cite{ma2016acoustic} in fields where vibrations play a crucial role, such as civil engineering and the aerospace industry.


\vspace{0.5cm}
\textbf{Acknowledgements}

M.M. has received funding from the European Union's Horizon 2020 research and innovation programme under the Marie Sk{\l}odowska-Curie grant agreement N. 658483. M. M. is also grateful to Dr. M. Mazzotti \& V. Pagneux for the insightful discussions.

\vspace{0.5cm}
\textbf{Author contribution}

All the authors contributed extensively to the work presented in this paper.

\vspace{0.5cm}
\textbf{Additional information}

Supplemental information is available in the online version of the paper. Correspondence and requests for material should be addressed to M.R. or M.M.

\vspace{0.5cm}
\textbf{Competing financial interests}

The authors declare no competing financial interests.

\vspace{0.5cm}
\textbf{Methods}

\small{\textbf{Simulations}. Dispersion diagrams and mode shapes presented in Figs. \ref{fig1} and \ref{fig2} are computed using Bloch-Floquet theory in full 3D FEM simulations carried out via the finite-element solver COMSOL Multiphysics. Full 3D models are implemented to capture all the possible wave modes. Linear elastic constitutive law is adopted and the following mechanical parameters for the waveguide considered: density $\rho = 2700 kg/m^3$, Young modulus $E = 70$ GPa, and Poisson ratio $\nu = 0.33$. Cells are meshed by means of 4-node tetrahedral elements of maximum size $L_{FE} = 0.5$ mm in order to provide accurate eigensolutions up to the maximum frequency of interest \cite{Miniaci_Ultrasonics_2014}. The band structures are derived assuming periodic (along the lattice vector $\textbf{a}_1$ and $\textbf{a}_2$ directions - see Fig. \ref{fig1}a) and free (in the out-of-plane direction) boundary conditions at the edges of the cell domains for all the cases considered in Fig. \ref{fig1}a. Dispersion diagrams shown in Fig. \ref{fig2} are computed instead considering $10 \times 1$ $x$-periodic strips. The resulting eigenvalue problem $(\mathbf{K}-\omega^2 \mathbf{M})\mathbf{u} = \mathbf{0}$ is solved by varying the wavevector $\vec{k} = \{k_x, k_y\}$ values along the boundary of the irreducible Brillouin zone $\Gamma - M - K$, with $\Gamma \equiv (0, 0)$, $M \equiv (0, 2/\sqrt{3})$ and $K \equiv (2/3, 2/\sqrt{3})$ for dispersion diagrams in Fig. \ref{fig1}b and $\vec{k} = \{k_x\}$ from $0$ to $\pi/a$ for band structures presented in Fig. \ref{fig2}. Colours in Fig. \ref{fig1}b indicate the mode polarization which is calculated defining a polarization factor $p=\frac{\int_V (|u_z |)^2 dV}{\int_V (|u_x |^2+|u_y |^2+ |u_z |^2)dV}$, where $V$ is the volume of the unit cell, $u_x$, $u_y$ and $u_z$ are the displacement components along $x$, $y$ and $z$ axes, respectively \cite{Miniaci_PRL_2017}. The dispersion curves are shaded accordingly, with colours varying from $0$ (blue) to $1$ (red), making the colour change from blue to red gradually. Thus colours close to red indicate vibration modes that are dominantly polarized out-of-plane, while the colours close to blue are predominantly polarized in-plane. The numerical wave-field reconstruction is performed via full 3D finite-element transient dynamic analysis in ABAQUS, where the experimental specimen geometry is accurately reproduced.

\vspace{0.5cm}
\textbf{Experimental measurements}. The topologically protected waveguide is fabricated through a two step machining process. First, the simple Kagome lattice is obtained through waterjet cutting, and then circular holes are drilled via computer assisted drilling process.
The waveguide consists of $20 \times 33$ unit cells. The Specimen is made of aluminium 6082 T6, with density $\rho = 2700 kg/m^3$, Young modulus $E = 70$ GPa, and Poisson ratio $\nu = 0.33$. The geometrical parameters are the following: $L = 20.5$ mm, $H = 5.9 $ mm, $h = 0.6$ mm, $R = 4$ mm, $r = 3.5$ mm.

Elastic waves are excited through a Krautkramer transducer ($1.25$ cm diameter and $1$ MHz central frequency) glued to the top surface of the waveguide and \MM{launching ultrasonic pulses made of 51 sine cycles Hanning windowed with a central frequency of $107$ kHz.} Refer to the SM for further details.

Dispersion curves for the edge modes (Figs. \ref{fig2}d and \ref{fig2}e) are obtained by 2D Fourier transforming \MM{the signals detected along the scanning line 1 and 2 reported in Fig. \ref{fig2}c. The specific positions allowed to follow a single mode (M1 or M2).} The maximum value of the energy as a function of the wavenumber and frequency is plotted.

The experimental wavefield reconstructions shown in Figs. \ref{fig2}f and \ref{fig4}a-c are obtained by means of scanning laser Doppler vibrometer measurements. \MM{The laser has been moved over ad-hoc designed grids due to the geometrical complexity. A step of approximately $0.2$ mm for the fine scan (Fig. \ref{fig2}f) and $1$ mm for the larger scan (Fig. \ref{fig4}a-c) have been used.}

Figs. \ref{fig4}e,f are obtained by means of a 2D-FFT performed on 1D-line scans of the same length token before and after the Z-bend. This allowed a quantitative evaluation of the lack of backscattering. Refer to the SM for the extended description of the experimental configurations.

In all the experiments no absorbers are placed at the boundaries of the samples to avoid any artificial dynamics alteration or mode conversion prevention during the propagation process.

\vspace{0.5cm}
\textbf{Data availability}. The data that support the plots within this paper and other findings of this study are available from the corresponding author upon request}.

\onecolumngrid

\end{document}